# ARCHITECTURE OF A IDENTITY BASED FIREWALL SYSTEM

Nenad Stojanovski[1] and Marjan Gušev[2]

1 Makedonski Telekom AD, Orce Nikolov BB, 1000 Skopje, Macedonia
nenad.stojanovski@telekom.mk
[2] Faculty of Natural Sciences and Mathematics, Ss. Cyril and Methodius University,
Arhimedova b.b., PO Box 162, 1000 Skopje, Macedonia
marjan@ii.edu.mk

*ABSTRACT*

*Classic firewall systems are built to filter traffic based on IP addresses, source and destination ports and protocol types. The modern networks have grown to a level where the possibility for users' mobility is a must. In such networks, modern firewalls may introduce such complexity where administration can become very frustrating since it needs the intervention of a firewall administrator. The solution for this problem is an identity based firewall system. In this paper we will present a new design of a firewall system that uses the user's identity to filter the traffic. In the design phase we will define key points which have to be satisfied as a crucial milestone for the functioning of the whole Identity based firewall system.*

## 1. INTRODUCTION

Classic firewall systems are built to filter traffic based on source and destination IP addresses, source and destination ports and protocol types[11]. As Information technology moves forward and advances, classic firewalls start to become very robust and unusable when it comes to the transparent user experience. The main problem that arises from using classic firewalls in a modern dynamic environment is that users have to be mobile and have access to their resources. The issue that is raised by classic firewalls is the one that forces the users to have static IP addresses, hence their needs for mobility[10]. This will require the firewall administrator to change the user's IP address in the firewall policy every time it changes. This may reflect on the user experience with the next generation network as it won't be as smooth as it should be, it might even become very frustrating for the end-user.

The solution for the problem is presented in the identity based firewall systems. This firewall solution should solve the problems that arise with user mobility so that it will allow users to be mobile in such way that there won't be any intervention needed from the firewall admin. Such solutions exist but are very robust because they need agents to be installed on client computers. This paper will focus on defining a new architecture and its components which will tend to solve the need of agents. The main focus of the solution will be to present a solution that won't be dependable on agents installed on the end-user computers. To do this, we need to build a very complex system which consolidates many different components that will aid the agentless system. In order to do this, we need to define this complex architecture for the identity based firewall and define the core components of that system. Along the way, we will define some crucial questions that have to be answered in the design phase of the solution. The main focus of this paper is to define the key points of the identity based firewall solution for an agentless solution. Also, we will give some results that we have acquired from the tests conducted on the proposed architecture.

## 2. USER IDENTITY IN THE COMPUTER WORLD

Like in the real world, the computer world has also needs for identities. These identities are needed to identify the computer users. To ensure this, system need to define schemas that will be used to identify users. These schemas are then used to map the user uniqueness in the computer world.

Microsoft's domain technology allows users to be defined in two different unique ways. Both ways are unique and both ways are mandatory for a user to be created as a domain user. One of those unique identifiers is called Unique User ID or short UUID[8]. The other unique identifier is the domain username. The first identifier is generated randomly by the active directory software, where the other one is inserted by an administrator.

The use of two unique identifiers for a user is very beneficial in an Active Directory environment. The most beneficial part is that computer systems are able to identify and distinguish user account. This means that by using the UUID, the operating system is able to check if the user is a domain user or if the user is a local user on that computer.

The other benefit of the Windows domain infrastructure is that every login and logoff operation is logged in a central repository. This is very important if there is a need for analyzing those events, the information that is included in them is very descriptive.

On the other side, the login process on the unix systems can be done in two different ways. The first possibility to solve the general question about user identities and logon procces is to use an LDAP which stores the login information[18]. This approach allows that the users are checked against the Microsoft Active Directory server. The second approach is the use of local users on the unix system. This approach doesn't support centralized authentication and is very hard for administration, since all users much be created locally on the unix systems[19].

## 3. EVENT DRIVEN SECURITY

Previously we've seen that almost every action is logged in the Windows domain environment. Depending on the definition in the group policy the Event is sometimes classified in a category or it's written as an error in the Event log files. By having the knowledge that everything is logged, it is possible to define a way how we will defend the systems based on the events that arise in those systems. By definition, event driven security is a security system that responds input from the user (mouse movement, keystrokes, menu choices, etc.) or from messages from other applications. This is in contrast to a batch operation that continuously processes the next item from a group.

By following the definition of event driven security, we can conclude that in order to have event driven security system, we have to monitor the events that arise in the monitored systems[7]. Since we know that every event is written in some log file we would have to monitor those log files for certain type of events. The main purpose of the event driven security system is to protect the assets by responding to malicious events. Because it is an event driven security system, when it detects any of the events that are specified as malicious it will execute the appropriate counter-measure to the destination systems.

For example, let's say that we have an Event Driven security defence system that protects the user accounts from being locked. That system will monitor the event log on the domain controller for events of the type "Event ID 529: Login failed". In order to protect it, the event driven system will start acting after the third failed login where it will send a command to the domain controller to block the IP address that had 3 failed logins.

Sometimes attacks are much more complicated and they don't appear straight forward as described in the example above. The anomaly that happens when they happen is connected with more than one event, and those events appear in different log files. In order to detect and defend against this type of attacks the event driven system should use event correlation. By using event correlation we are able to define sequential list of events that occur during some attacks. This technique will allow us to stop the attack before the last step of the attack occurs.

For example, the attack against our system happens in three steps where the steps are:

1. login to the operating system
2. warning in the system log that service X is failing
3. error in the system log that service X has stopped responding

In order for the event driven system to protect against this type of attack we should define a 2 step event correlation[22] where the rule will be:

1. login to the system
2. warning in the system log that service X is failing after x seconds from the login event

When this rule is detected by the event driven system it should send an action to the attacked system to block the access from the attacking IP address. Other uses of the event driven security paradigm can be that we can define actions that will occur when certain events are written to the event log.

## 4. IDENTITY BASED FIREWALL

In the last few years the Microsoft Windows domain infrastructure has spread rapidly among companies. This has brought the ability to uniquely identify users in this domain, since every user has been given a unique domain user name. Moreover, with the rapid development of new devices, like smart phones, laptops and the lowering of the prices for them, the need of user mobility within these the modern enterprise rises. The whole concept of mobility asks that the users are mobile within their work environment, which forces them to use different IP addresses. By using different IP addresses, they are limited in the access to their resources. In order to solve the issues with the mobility, security companies have started to design and develop new concepts that could be implemented into the current firewall technology. These concepts should allow users to extend their mobility and let them be mobile in such ways that they would be able to access resources from where ever they are.

The whole concept of identity based firewall systems differs from the classic packet filtering solutions in many ways. The main difference is that identity based firewall systems have to use the layer 7 information to do the filtering, where the classic solutions use the layer 3 information. The whole idea behind the identity based firewall concept is to use the user's identity and allowing that identity access to the needed resources. In a Microsoft Windows domain environment the main identity is the user's domain username. Rules in the case of identity based firewalls are created by assigning permit or deny permissions to user identities. This means that we are embedding the user identity within the standard firewall rules, but instead of using a source IP address, the rule will use the user identity. In the case of the windows domain, the user identity is represented with a domain username. By using this domain username and adding it to the identity rules, the windows domain users will be able to access their resources with the use of their domain username.

## 5. IDENTITY BASED FIREWALL ARCHITECTURE

Identity based firewalls introduce much bigger complexity in the computer network then classic firewall. The complexity of this new architecture is introduced by the use of the Layer 7 information according to which the firewall filtering is done. Namely, the complexity is brought to the network by the following requirements:

1. A system that will collect the layer 7 data (username) from the central authentication device
2. Agentless system when clients are concerned
3. Optimized system where the introduced identity will have a minimal performance on the user's experience

The previous requirements are crucial in building a very optimized and very functional identity based firewall[5]. By fulfilling every requirement from above we will be able to define a very optimized identity based firewall.

In order to answer the previous requirements we will first describe the "as-is" process of the user's access to the network and his/hers access to the needed resources that are in the network. The scenario is based on the assumption that the users are mobile users and that they use IP addresses that are being dynamically assigned to the computer and all network resources are in a secure network zone which is behind a firewall. First step in the whole process is the need for the user to log into the computer. In background, the operating system authenticates the user to a central system. After being authenticated to the system the user wants to access the resources in the network. Since the user is a mobile user it is very likely that there will be a problem in the access to the resources. The reason behind this is that the user is being blocked by the firewall, because his dynamically IP address isn't allowed to the resources. This is true because classic firewalls require that the user is defined by an IP address. Since the IP address is dynamically assigned, the firewall admin is required to change the access rules for every user that has a new IP address assignment.

After analyzing the previous scenario we can highlight couple of points in the user's activity towards his access to the network resources. We can distinguish three main points which are:

1. The user's authentication to the central authentication system
2. Reporting the new issue to the firewall admin
3. The reconfiguration needed on the firewall with the user's new IP address

The following scenario will give a new overview with the implementation of the new identity based firewall with the use of the event driven security paradigm. Our main goal is to build an identity based firewall solution and to do this we will need to redesign the previous scenario to fit for our identity based firewall. This means that the following steps need to be done:

1. Catch every logon/logoff event on the domain controller
2. Translate the Layer 7 information into Layer 3 information
3. Recompile the firewall policy with the changes caused by the logon/logoff event
4. Install the policy on the firewall

By analyzing the previously defined 4 steps we can derive 4 crucial questions that will define the architecture of the Identity based firewall system in details. The four questions that need to be answered are the following:

1. How can we know when someone logged to the domain and on which computer did he logged in?
2. How can we convert the Layer 7 data (username) into something easily filtered (IP address)?
3. How can we build a firewall policy that will be template like and that will support multiple reconfigurations in a small timeframe?
4. How can we create a readable policy for the firewall by using the policy template?

The next chapter will concentrate in giving answers to these questions and it will define the systems architecture in detail.
In order to comply with the previous 4 steps we will need to introduce four components to the classic firewall and extend it to an identity based firewall. The needed components are: a sensor, a translator, a compiler and an installer component.

## 6. IDENTITY BASED FIREWALL ARCHITECTURE IN DETAILS

We mentioned before that the key of defining the architecture for the Identity based firewall is to answer the questions that concern that system. This chapter will tend to answer those questions in detail as well as give detailed information about the Identity based firewall architecture and its components. The answer to the first question can be found in the Windows Event Log that is generated on the domain controllers. These events are logged into the Security Event Log of the Windows operating system. The information represented in these event logs is very easily extracted since all the crucial data is written in a form of an event message. Login and Logoff events are represented with the following event IDs:

- 540 is the ID of the Login event [7]
- 538 is the ID of the Logoff event [7]

The answer to the second question is in the Event Logs[20]. Namely, the event logs contain the information of the IP address that was used by the user to Login to the Windows Domain. By extracting the username and IP address from the Event Log we have the valid conversion of the Layer 7 information into a Layer 3 format. If the systems are non Microsoft Windows and the login authentication is not done by using the Microsoft Windows Active Directory, then those systems can be also integrated in the Identity firewall. This is done by using the syslog functionality on the Unix systems in such way that all authentication data is sent in a form of a syslog message[7][18] to the Identity firewall. The syslog messages will contain the username that logged to the system and the IP address of the syslog message sender will be the Layer3 information regarding that username[21]. By knowing the answers to the first two questions we introduce a new component to the Identity Based firewall architecture which we will name "IdentityAgent". This Identity agent will have the role to read the Event Logs from the domain controllers, extract the data from the event logs and send all the relevant updates to another component that will use that data to trigger some action. The IdentityAgent will read the Event Log data via the Windows Management Instrumentation – WMI[8]. The whole WMI concept enables us to create a fully agentless system, since all components will be on the identity firewall. In this way we can read the events in real-time and we can send updates to the other component in real-time. By using this methodology we introduce a new way of catching user logins without having the need to install agent on all client computers.
The answer to the third question lies in the use of XML files to built policy templates. These XML files will hold user based meta policies. In the meta policies, the rules will be based upon a domain username. In this way, every change of the user's IP address will result in compilation of the meta policies with the new changes that happened in the Domain. At this point we will introduce a second component of the Identity Based firewall system. We will name this component "IdentityCompiler". The primary function of this component will be to receive updates from the first component and after receiving new updates, the IdentityCompiler will scan through the meta policy and make the appropriate changes for that user. After the changes are made, the IdentityCompiler will compile the meta policy in a firewall readable form. Since the IdentityCompiler creates a firewall readable policy by following the compilation rules we can conclude that the fourth question is also answered and that we can create a reliable method for compilation. By now, we successfully implemented the event driven security paradigm [4] in our Identity based firewall architecture.

On the other hand, by using this way of handling the user traffic we create a very limited system that can only handle authenticated traffic. In order to resolve this limitation, we adopted the system to allow also addition of layer 3 rules. By using this approach we can handle guest access and access through mobile phones and tablets.

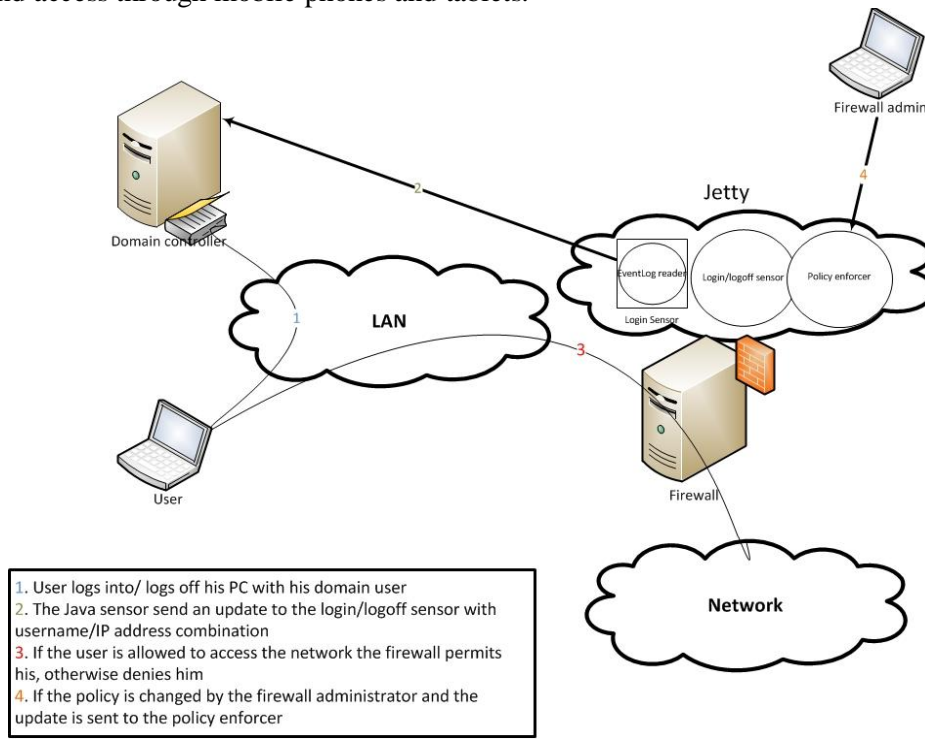

Figure 1. Identity based firewall system architecture

## 7. TESTING RESULTS

In the process of testing we tested the time needed for the user policy to become active. The tests were done with 10, 15, 20, 25 and 30 clients. Mainly, the tests were focused on measuring the time needed for the new client detection and installation of the policy for that client. After we collected the measures we calculated the average value for every solution. We calculated that the average time for installation of a policy with a classical firewall solution is 120 seconds, where the agent based identity firewall solution had an average of 7 seconds. This measurement is a sum from the time needed for the user to log on the domain and the time needed for the local agent to send the update to the identity based firewall, since the agent has to be started under the logged on user context. Our proposed solution had an average time of 5 seconds, because it detects the logged on user right away upon authentication on the domain. In the end, we multiplied the average values with the number of clients used in the test assuming that all of the events occurred at once and all of the policy installation is done one by one for every client. Figure 2 and Figure 3 represent the charts that we created from the tests.

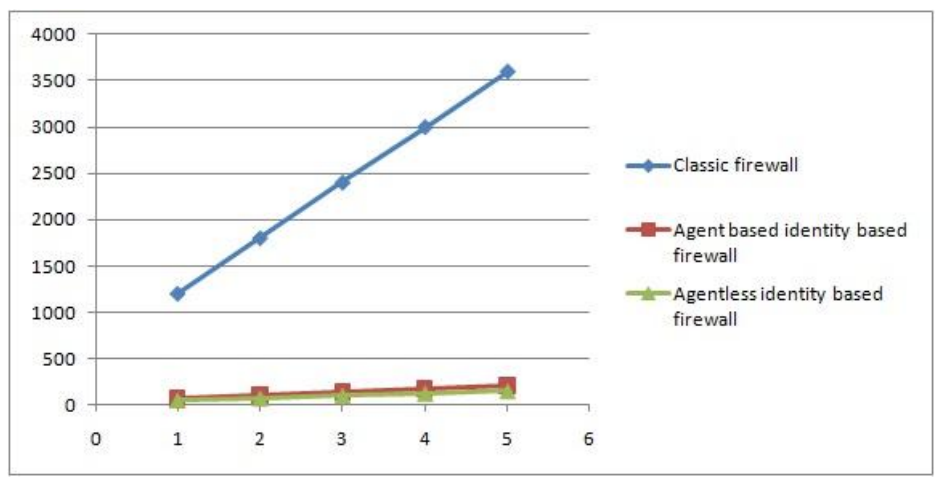

Figure 2. Charts that include the classic firewall solution in the tests

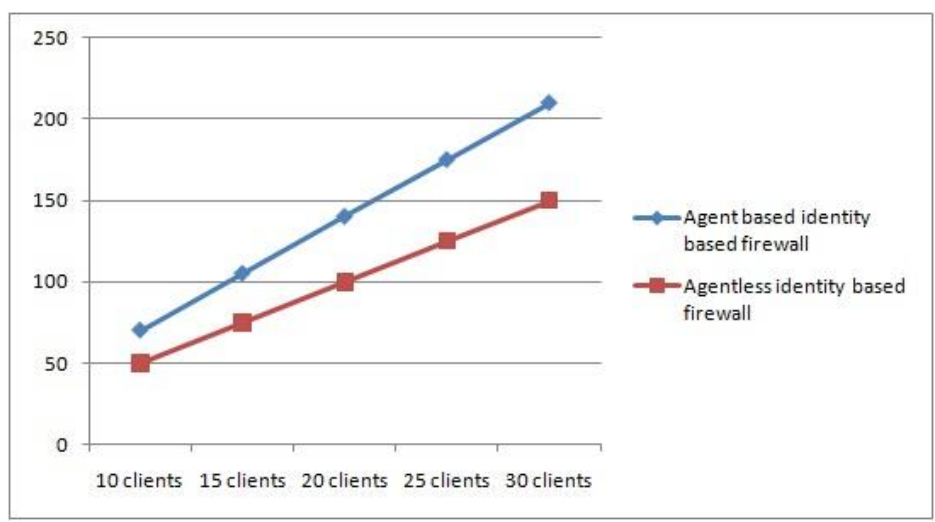

Figure 3. Charts without the classic firewall

## 8. ARCHITECTURE PROS AND CONS

As one of the key benefits of our proposed architecture is that there is no need of agents on the client operating systems. In this way, the presented design minimizes the network overhead on all of the nodes and concentrates that overhead on single points, the domain controllers. In comparison to other solutions[6], our architecture tends to trigger login or logoff updates only once(when the event occurs), whereas other solutions send those updates periodically[20].

Other benefit of this design is the ease of management. Other solutions add some administrative effort in management, since every computer has its own agent. Also, as a big benefit we can point out the ease of troubleshooting and minimization of downtime, since agents that read the event logs are located on the Identity firewall.

As a con to this solution we can point out that when the client is powered of by force and when the domain controller doesn't receive a log off event, and then the policy isn't updated to do a cleanup of the installed policy.

It has to be mentioned that one con for all identity based firewalls is that they don't support multiuser systems like Microsoft Terminal Services or Citrix, because that feature would require that the user information is sent in every packet.

Table 1 represents the comparison between our architecture and other solutions. It has to be noted that Microsoft ISA introduces limited firewall filtering without agents (only filtering of web traffic).

Table 1. Comparison chart between the analyzed solutions and our architecture.

|  | **Microsoft ISA 2006** | **NuFW** | **CheckPoint UserAuthority** | **Proposed Architecture** |
|---|---|---|---|---|
| L3, L4 filtering | ✓ | ✓ | ✓ | ✓ |
| Identity based filtering without agents | ✓ * | ✗ | ✗ | ✓ |
| Identity based filtering with agents | ✓ | ✓ | ✓ | ✗ |
| Network overhead | ✗ | ✓ | ✗ | ✓ |
| Multiple OS Support | ✗ | ✓ | ✗ | ✓ |

## CONCLUSION

The research presented in this paper develops an innovative approach for identity based firewall architecture. Specifically, it presents an architecture that will enforce identity based firewall filtering without the need of having an agent installed on every computer that participates in the communication with resources located behind the identity based firewall. Throughout a case study conducted on a set of users, the architecture can be proven as a very effective.

The research has several important contributions. First, the approach is simple and it uses existing resources that exist in the infrastructure to do the job. Second, the core of the identity based firewall is build upon open source technologies, which are extended to exploit existing proprietary technologies. Third, the architecture and its core are very flexible, and it can be extended with new components. On the other side, this approach minimizes the points of failure and pinpoints then on one place, the firewall itself. Finally, the biggest benefit is the minimization of the network overhead and minimization of the traffic cost that goes through the wire. By making this possible, the low overhead can be valued either as a direct or indirect benefit. Overall, the presented architecture in this research proved that there are benefits with the use of the agentless identity based firewall system. Moreover, it also proves that it is possible to implement the solutions for multiplatform environments.

## ACKNOWLEDGEMENTS

The authors would like to thank the reviewers whose constructive critique greatly improved the quality of the paper.

## REFERENCES

[1] Amon, C., Shinder, T., Carasik-Henmi, A.: The Best Damn Firewall Book Period, Syngress, 2003

[2] Gizem, Aksahya & Ayese, Ozcan (2009) *Coomunications & Networks*, Network Books, ABC Publishers.

[3] Noonan, W., Dubrawsky, I.:Firewall Fundamentals, Cisco Press (2006)

[4] Zwicky, E., Cooper, S., Chapman, D.:Building Internet Firewalls, O'Reilly Media Inc. (2000)